\documentclass[british,english,a4paper,superscriptaddress,aps,prl,notitlepage,twocolumn,showkeys,showpacs]{revtex4-1}

\usepackage{graphicx}
\usepackage{hyperref}
\usepackage{amsmath}

\newcommand{\ree}{\langle R^2_{\rm ee}\rangle}
\begin{document}


\title{Mapping onto ideal chains profoundly overestimates self-entanglements in polymer melts}
\title{Mapping onto ideal chains overestimates self-entanglements in polymer melts}

\author{H. \surname{Meyer}}
\email{hendrik.meyer@ics-cnrs.unistra.fr}
\affiliation{Institut Charles Sadron, Universit\'e de Strasbourg, CNRS UPR 22, 23 rue du Loess-BP 84047, 67034 Strasbourg Cedex 2, France}
\author{E. \surname{Horwath}}
\author{P. \surname{Virnau}}
\email{virnau@uni-mainz.de}
\affiliation{Institut f\"ur Physik, Johannes Gutenberg-Universit\"at Mainz, Staudinger Weg 7, 55099 Mainz, Germany}

\date{\today}
           

\begin{abstract} 
In polymer physics it is typically assumed that excluded volume interactions are effectively 
screened in polymer melts. Hence, chains could be described by an effective random walk without excluded 
volume interactions. In this letter, we show that this mapping is problematic by analyzing the occurrence of 
knots, their spectrum and sizes in polymer melts, corresponding random walks and chains in dilute solution. 
The effective random walk severely overrates the occurrence of knots and their complexity, particularly when compared to melts of flexible chains, indicating that 
non-trivial effects due to remnants of self-avoidance still play a significant role for the chain lengths 
considered in this numerical study. For melts of semiflexible chains, the effect is less pronounced.
In addition, we find that chains in a melt are very similar in structure and topology to dilute single chains close to
the collapse transition, which indicates that the latter are also not well-represented by random walks.
We finally show that typical equilibration procedures are well-suited to relax the topology in melts.

\end{abstract}




\maketitle

A first link between physics and knots was established when in 1867 Lord Kelvin speculated \cite{thompson} 
that atoms and molecules may be composed of knots in the ether. Even though this beautiful hypothesis was 
eventually rejected, arguably, it ushered in the era of modern mathematical knot theory. About a hundred years 
later, knots were rediscovered by natural scientists, stimulated and encouraged by theoretical considerations 
about knotted polymers \cite{frisch,delbrueck} and somewhat later by the discovery of knotted DNA 
\cite{macgrego,liuDepew,dean,arsuaga02Knots,arsuaga05,Virnau_chromosomes} and proteins 
\cite{mansfieldKnotsProteins,takusagawaRealKnot,taylorDeeply,Jackson_05,virnauIntricate,Grosberg_proteins,Sulkowska_09,boelinger,virnauStructures}.

In polymer science the investigation of self-entanglements and knots can be traced back to the so-called 
Frisch-Wasserman-Delbruck conjecture from the early 1960s \cite{frisch,delbrueck}. This conjecture essentially 
states that all polymers will eventually be knotted as the chain length of the polymer increases which coincides 
with our observations of macroscopic ropes and strings. Not only did these early works stimulate the synthesis 
of knotted polymer rings \cite{dietrichbuchecker,forgan}, they also led to numerous investigations in 
the context of statistical physics, e.g., 
\cite{vologodskiiKnot,frankkamenetskii,vanrensburg,mansfieldHamilton,grosbergCrit,virnau05PE,dobay,rosaJamming,trefz,Virnau_dna_knots,doyle_15,Janke_16}. 
Some basic classes of single polymers have already been considered in the first simulations of knotted 
polymers \cite{vologodskiiKnot,frankkamenetskii}: Random walks (RW) tend to be highly 
knotted as it is easy to form small local loops without excluded volume interactions. In contrast, polymers with excluded volume interactions tend to knot 
at considerably larger chain lengths as repulsion between monomers inhibits the formation of local loops 
\cite{koniarisKnottedness,virnau05PE}. Globular polymers and polymers in confinement, however, tend to be rather knotted again 
\cite{mansfieldHamilton,virnau05PE,Micheletti_polymer_review}. Even though numerous studies exist on single polymers, little is known about knots 
in concentrated solutions or melts. Only in \cite{foteinopoulou,lasoRandom}, knotting probabilities in a melt of tethered 
hard sphere chains were determined to depict connections between intra- and interchain-entanglements. In \cite{sommer_knots} knotting probabilities of
ring polymers in semidilute solutions were investigated.

\begin{figure}[htb]
	\centering
	\includegraphics[width=0.9\linewidth]{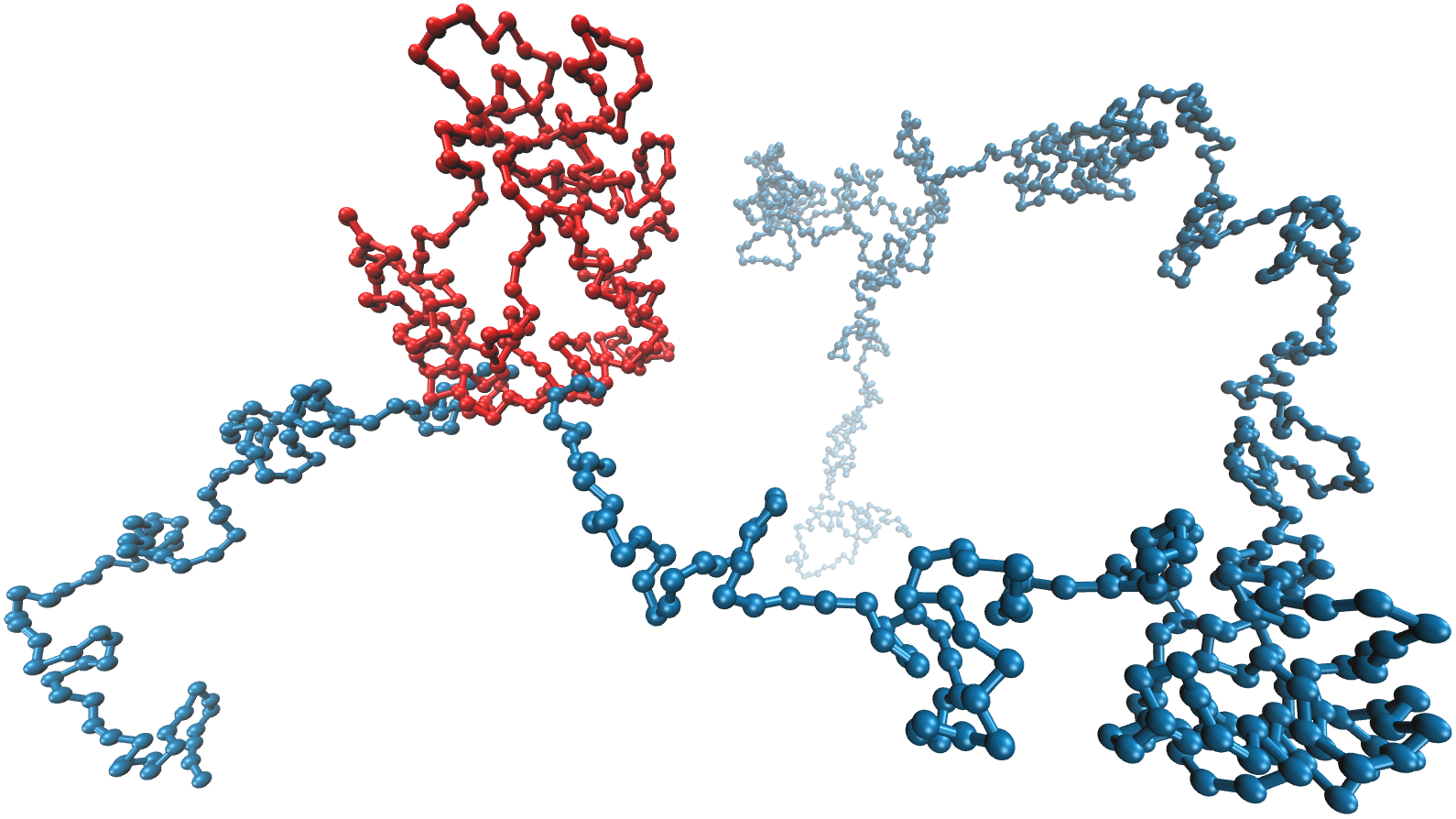}
	\caption{Single chain ($N=1024$) from the polymer melt with $3_1$ knot of typical size and end-to-end distance rendered by VMD\cite{vmd}. The trefoil knot is highlighted in red.}
	\label{fig:vmd_pic}
\end{figure}

From a mathematical point of view, knots are only well-defined in closed loops. Therefore, the termini of open chains need to 
be connected in a well-defined manner before the knot type is determined. Even though the closure can in principle lead to the 
creation of additional entanglements and knots, knotting probabilities for different closures vary only little \cite{virnau05PE,Detection}. 
In this work we apply a closure which has been used successfully in the context of protein knots \cite{virnauIntricate}: We 
draw and connect two segments starting from the end monomers to the outside of the polymer along connection lines between the 
center of mass and the respective terminus. Knots are then detected with a variant of the Alexander polynomial as described in 
detail in reference \cite{Detection}. We also characterize the size of the knot by successively deleting bonds first from one 
side, then from the other side until the Alexander polynomial changes \cite{virnau05PE}.

Since the work of Flory\cite{Flo1969book}, chains in the melt are supposed to be ideal because of screening of excluded volume interactions. 
The mean-square size of a polymer of $N$ monomers is thus described by
\begin{equation}
 \langle R^2\rangle = l^2 C_\infty N = a^2 N^\star
 \label{defree}
\end{equation}
with the bond length $l$, the so-called Flory characteristic ratio $C_\infty$, and $a$ being the statistical segment length of the equivalent RW with $N^\star$ steps.
With the constraint of the contour length $lN=aN^\star$ one gets $N^\star=N/C_\infty$.
The chemical details determine the characteristic ratio. For lengths longer than $a=lC_\infty$, the chain is considered as a RW.
Angular correlation between nearest neighbor beads 
$c=\langle \cos\theta\rangle=\langle \vec{u}_i \cdot \vec{u}_{i+1}\rangle/l^2$
are included in the freely rotating chain (FRC) model which describes the approach to the RW behavior for finite chain lengths or sub chains of size $s$:
\begin{equation}
 \langle R^2(s)\rangle = sl^2 \left[ C_\infty  - \frac{2c(1-c)^s}{(1-c)^2} \frac{1}{s}\right]
 \label{frc}
\end{equation}
with $C_\infty=(1+c)/(1-c)$.
Additional terms have been calculated if the torsional distribution is not flat \cite{Flo1969book}.
Although this gives reasonable estimates of the size of large polymer chains for chemically realistic models,
Eq.~(\ref{frc}) (by using the local $c$) severely underestimates the chain size for flexible polymer models.

It was discovered recently that screening of excluded volume interactions is actually not complete in polymer melts and that long-range interactions lead to systematic corrections to scaling
\cite{WiMeBaEA2004prl,WiBeJoEA2007epl,WiBeMeEA2007pre,BeJoSeEA2007macro,MeWiKrEA2008epje,WiEA2011review,GrQiMo2007pre}; 
\begin{equation}
  \langle R^2(s)\rangle = sb^2 \left[ 1- \frac{\sqrt{24/\pi^3}}{\rho b^3}  \frac{1}{\sqrt{s}} \right]
  \label{r2s}
\end{equation}
where $b$ is a renormalized statistical segment length and $\rho$ the monomer density. 
A signature of the corrections to ideality is a power-law decay of the bond-bond correlation along the chain $P_1(s)=\langle \vec{u}_i \cdot \vec{u}_{i+s}\rangle/l^2 \sim 1/s^{3/2}$.
Similar corrections have been found to hold for dilute chains at theta conditions \cite{ShPaLiRu2008macro}.
Note that the approach to the asymptotic RW behavior with the term $1/\sqrt{s}$  is much slower
compared to the $1/s$ correction in the FRC model Eq.~(\ref{frc}). 
However, the amplitude of the correction in Eq.~(\ref{r2s}) decreases with $1/b^3$ which means that this correction becomes rapidly weaker with increasing persistence length.
Although in principle always present \cite[fig4]{albert-2dreview}, 
the effect of the long-range corrections is thus weak as soon as $C_\infty\gtrsim 3$ and 
for example the form factor is dominated by the rigidity instead of the corrections to ideality \cite{HsKr2016jcp}; synthetic polymers have typically $C_\infty=4\dots7$.

In addition to the size of (sub)chains \cite{WiBeMeEA2007pre}, knots are a fine gauge for structural properties of polymer chains. 
In this work we compare knottedness, knot complexity and sizes of typical knots occurring in polymer melts (see Fig.~\ref{fig:vmd_pic}) with the same properties 
in equivalent RWs. We find that, contrary to general belief, RWs provide a rather poor description of 
the structure of chains in a melt and tend to overemphasize the role of self-entanglements. We conjecture that this failure 
can be attributed to the incomplete screening of self-avoidance at the local scale.


The melt configurations analyzed in this work have been obtained by standard molecular dynamics (MD) simulations using the code LAMMPS \cite{Pli1995jcp}.
The polymer model is a generic bead-spring model with purely repulsive Lennard-Jones (LJ) beads
\begin{equation}
V_{\rm LJ}^{\rm rc}(r)=\left\{\begin{array}{ll}
             V_{\rm LJ}(r) - V_{\rm LJ}(r_c)& \mbox{, if\hspace{0.3cm} $r<r_c=\sqrt[6]{2}$$\sigma$} \\
             0 & \mbox{, else.}
             \end{array} 
       \right. 
\label{LJshift}
\end{equation}
with $V_{\rm LJ}(r)=4\epsilon\left[\left(\frac{\sigma}{r}\right)^{12}-\left(\frac{\sigma}{r}\right)^{6}\right]$ 
and connected by harmonic springs
$ V_b(r_{i,i+1}) = 400 (r_{i,i+1}-0.967)^2$.
The parameters of the harmonic bonds are chosen such that the average bond length is the same as for the frequently used FENE potential \cite{Auhl2003jcp}.
We use LJ units with $\sigma=1$, $\epsilon=1$, $T=1$.
To vary the persistence length, an angular potential $V_a(\theta)=B(1-\cos\theta)$ has been added with prefactors $B=0$ (completely flexible), $B=1,2,4$.
Polymer melts have been generated from FRCs with the expected Flory ratio $C_\infty$, similar to the procedure described by Auhl {\it et al.} \cite{Auhl2003jcp}.
The chains have been randomly moved in the box for a short time to pre-equilibrate the density before
adding the LJ potential gradually  by increasing a force-cap parameter \cite{Auhl2003jcp,MeWiKrEA2008epje}.
A large system with $N=1024$ and 768 chains has been generated at monomer density $\rho=0.68$ and run for more than five Rouse times during which the end-to-end vector autocorrelation has decayed to 1\%.
Shorter chain systems have been derived by cutting the longer chains and allowing for additional equilibration, or by independent setup. In a similar manner, systems have been prepared at different densities.

\begin{figure}[tb]
	\centering
    \includegraphics[width=0.95\linewidth]{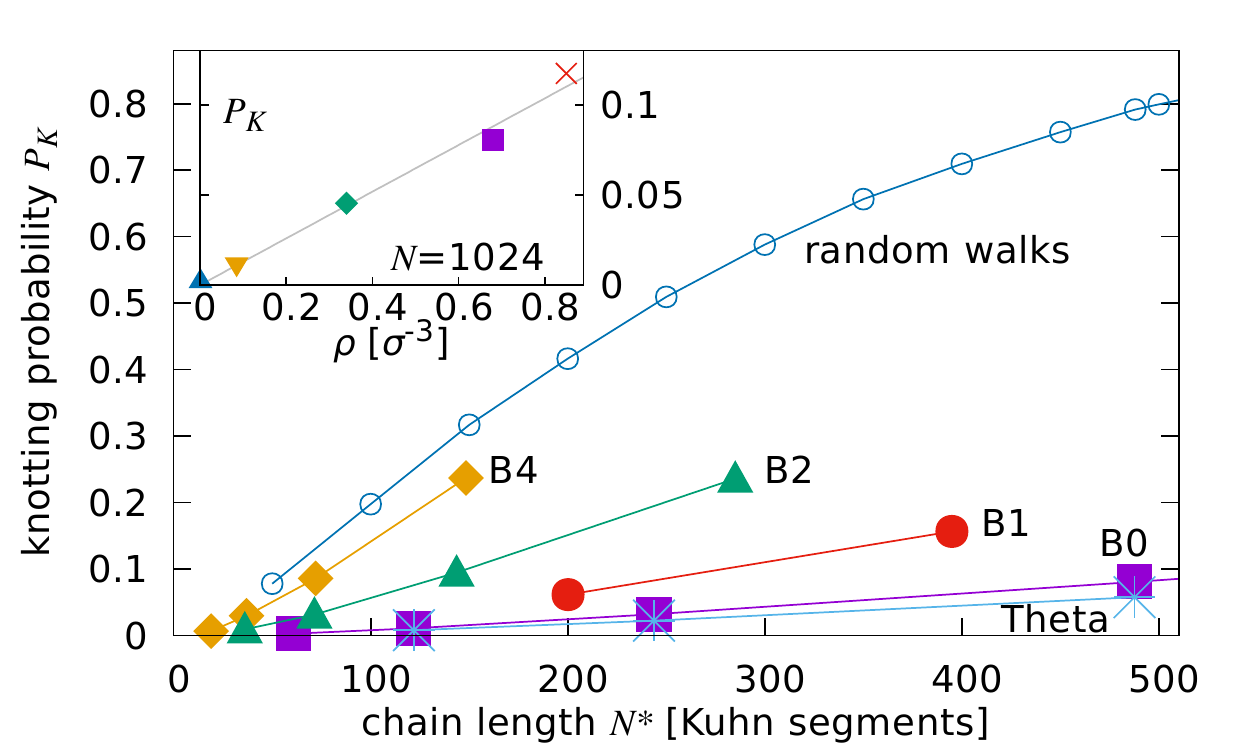}
    
\caption{Knotting probability for RWs (open circles) compared to chains in polymer melt with a density $\rho = 0.68\sigma^{-3}$ (filled symbols)
  with different angular potential (B0=completely flexible). Results for $\Theta$-chains of size equivalent to B0-chains are shown by stars.
  The chain length of the different models has been converted to $N^\star=N/C_\infty$, the number of Kuhn segments, to compare with the equivalent RW
  $C_\infty^{\rm B0}=2.1$, $C_\infty^{\rm B1}=2.6$, $C_\infty^{\rm B2}=3.6$, $C_\infty^{\rm B4}=7$).
 The inset shows higher and lower densities of the flexible model (B0) for constant chain length $N=1024$, the grey line is a guide to the eye. 
 The error is smaller than the symbol size.
}
\label{fig:knot_paper}
\end{figure}

\begin{figure}[tb]
        \centering
 \includegraphics[width=0.95\linewidth]{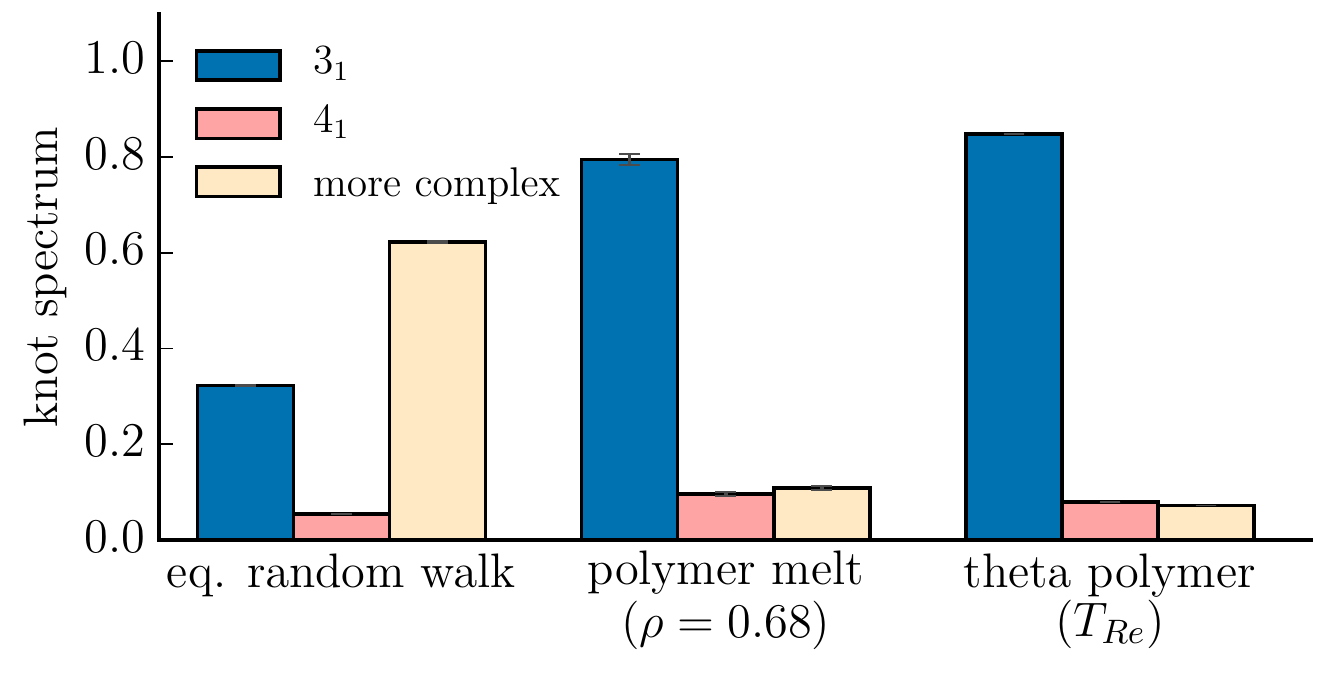}
 \caption{Knot spectrum of the two simplest knots $3_1$ and $4_1$ from chains of the length $N=1024$ 
        (RW with chain length $N^*=N/C_{\infty}^{(\rho=0.68)}=488$, equivalent to a polymer chain length $N=1024$ with $C_{\infty}^{(\rho=0.68)} = 2.1$). }
 \label{fig:knotSpectrum}
\end{figure}

\begin{figure}[htb]
        \centering
  \includegraphics[width=0.95\linewidth]{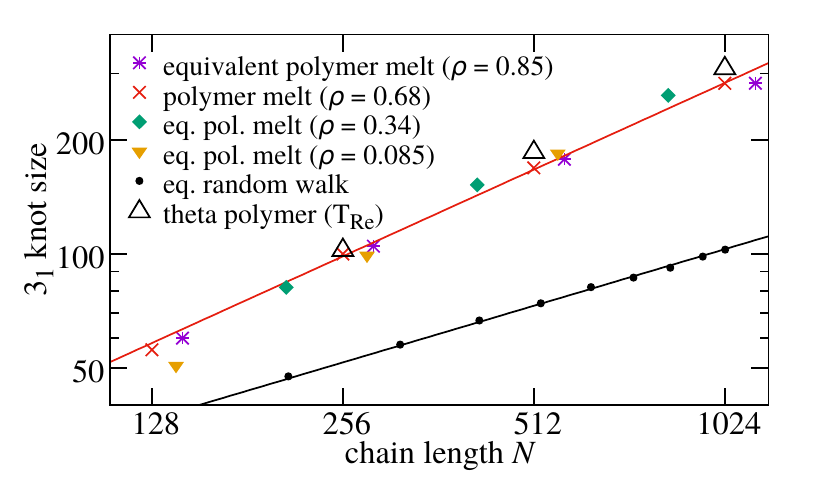}
  \caption{Knot size of $3_1$ knot from chains in a polymer melt with different density compared with RW. 
     The chain length $N^*$ and knot size $S^*(3_1)$ of all polymer melts and RWs are rescaled to a same $C_{\infty}^{(\rho=0.68)}=2.1\,$. 
     Equivalent chain lengths $N^* $ for chains in polymer melts/solutions are calculated by $N^*=N \cdot 2.1/C_{\infty}$. 
     Likewise the equivalent knot size $S^*(3_1) = S(3_1) \cdot 2.1/C_{\infty}$. 
     Straight lines are power-law fits to the data with errors smaller than the symbol size.} 
  \label{fig:knotSize31}
\end{figure}

\begin{figure}[htb]
    \centering
    \includegraphics[width=0.95\linewidth]{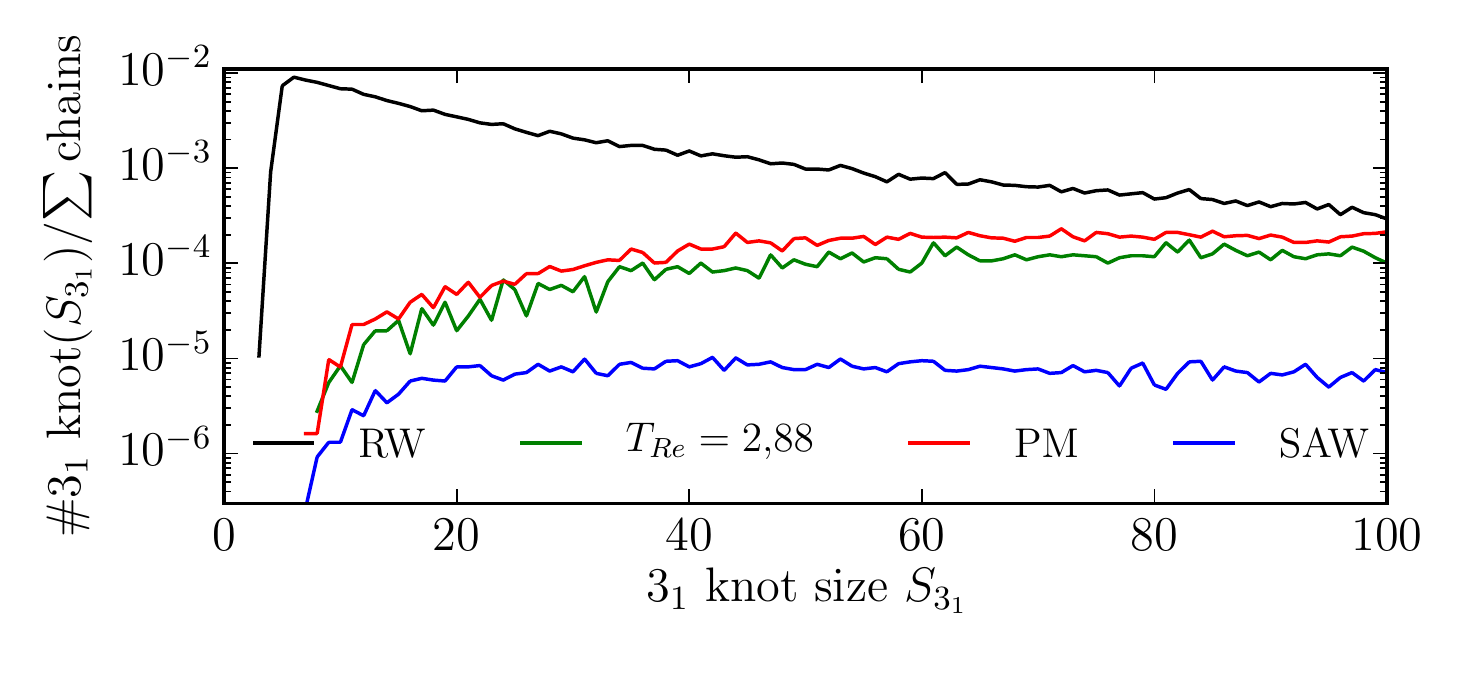}
    \caption{Probability of observing a trefoil knot of a particular size. Chains
in a melt ($\rho=0.68$, $N=1024$, B0) are compared with the distribution of a $\Theta$-like chain with length $N=1024$ and the same $\ree$ as the chain in the melt. For comparison, we have also plotted
the distributions for RWs and self-avoiding walks of size $N=500$.}
    \label{fig:knotSizeDistribution}
\end{figure}

Our main result is shown in Fig.~\ref{fig:knot_paper}, which compares the probability of observing a knot 
in a single polymer $P_K$ extracted from a melt at density $\rho=0.68\ \sigma^{-3}$ with the knotting 
probability in a corresponding RW of length $N^*=N/C_{\infty}$ as a function of chain length. 
The corresponding RWs severely overrate the occurrence of knots, e.g. for $N=1024$ ($N^\star=488$), we measure a 
knotting probability of roughly 8\%, whereas knots occur in $80\%$ of all configurations in RWs. 
For a dilute self-avoiding polymer, $P_K$ is almost zero for the 
chain lengths considered (e.g., $P_K=0.008$ for the leftmost point of the inset Fig.~\ref{fig:knot_paper}). We have verified that knots form and unform over the time evolution,
but the average knotting probability does not change over time.
Fig.~\ref{fig:knot_paper} also shows data for melt chains with angular potential and thus higher $C_\infty$.
In this case the knotting probability at equivalent $N^\star$ increases, but still remains much smaller than for the equivalent RW.

Differences can also be identified when the whole spectrum of knots is taken into account. 
In the following, we restrict the discussion to the flexible model (B0) for which the effects are most pronounced.
Fig.~\ref{fig:knotSpectrum} shows the probabilities of observing a trefoil ($3_1$), a figure-eight knot ($4_1$) 
and more complex knots for $N=1024$ and $N^*=N/C_{\infty}$. Knots in corresponding RWs are 
typically more complex whereas the spectrum for a chain in the melt is still dominated by trefoil knots similar 
to the spectrum in simple self-avoiding polymers. 
Fig.~\ref{fig:knotSize31} shows the average size  of a trefoil 
knot as a function of the chain length normalized by $C_{\infty}$ for $\rho=0.68\ \sigma^{-3}\,$. 
The average size of a knot in a corresponding RW is much smaller and more localized than a typical knot in a melt. 
Interestingly, data for knot sizes taken at different polymer densities are all compatible with each other once they are properly normalized.
Fig.~\ref{fig:knotSizeDistribution} combines the information of Figs.~\ref{fig:knot_paper} and \ref{fig:knotSize31} 
and shows the distribution of the probability to obtain a trefoil knot of a certain size. The integral over 
this curve yields the overall probability to observe a trefoil knot in a configuration. At the
same time the plot provides information about the knot size distribution, which goes beyond the average size displayed 
in Fig.~\ref{fig:knotSize31}. It is therefore a fine gauge to analyse the overall structure of a chain in a melt.
In conclusion, the RW overpredicts the occurrence of knots and underestimates their size.

To rationalize the findings of Fig.~\ref{fig:knot_paper}--\ref{fig:knotSizeDistribution}, 
consider the bond-bond correlations $P_1(s)$ along a chain which are zero for a pure RW and decay exponentially in the FRC model.
To form a knot, the chain needs to come back close-by to wind around itself, which represents a negative bond-bond correlation on the scale of the knot.
For real chains, the impossibility to occupy the same place twice generates the corrections to ideality with (on average) a strictly positive power-law tail in $P_1(s)$ \cite{WiMeBaEA2004prl}. 
A knot thus needs an unfavorable fluctuation with respect to the  average.
As bond-bond correlations are stronger at short distances, small knots are suppressed more strongly than larger ones.
With an internal bending potential, the effect of the bond-bond correlations is weakened which explains that the results with larger $B$ have higher knotting probability at equivalent $N^\star$.

In search for a better representation of the structure of chains in a melt via a single chain model, we have tried several approaches: 
Reducing excluded volume interactions by either diminishing the size of beads as in \cite{koniarisKnottedness,koniarisEntanglement} or restricting interactions 
to neighboring beads (finite memory walker) \cite{Horwath2016FMW} did not provide satisfactory results even though both are in 
principle able to reduce knotting. 
Amongst others, it was impossible to match the distribution of trefoil knot sizes (Fig.~\ref{fig:knotSizeDistribution})
and small knots were typically overrepresented similar to a RW, even though the overall probability is smaller 
(not shown). The finite memory walker in particular closely followed the distribution 
of self-avoiding walks up to the number of neighbors included for the excluded volume interactions, 
before following the distribution of RWs for larger distances \cite{Horwath2016FMW}. 

A good match is, however, provided by 
simulation of single chains in a solvent close to its finite-size $\Theta$-point. To this extent we have modified the 
model in Eq.~(\ref{LJshift}) by using as cutoff $r_c=2\sqrt[6]{2}$ to include attractions.
In this case, the FENE interaction was used for the connectivity
$V_{\rm FENE}=-33.75\epsilon\cdot \ln\left[1-\left(\frac{r}{1.5\sigma}\right)^2\right]$.
For large temperatures, attractions between beads are suppressed and the chain will resemble a self-avoiding walk. 
For low temperatures, energetic contributions to the free energy of the polymer will dominate and the chain will 
collapse to a globular conformation. At the (size-dependent) transition temperature, the chain will neither be 
swollen nor globular and adapts a conformation which is again asymptotically described by a RW \cite{ShPaLiRu2008macro}. Here, we 
take a more pragmatic approach. The temperature (for each $N$) was chosen so that the quadratic 
end-to-end distance $\ree$ of our single chains and the chains in the melt ($\rho=0.68$) of the same size match 
($T=2.81(1), 2.85(1)$ and $2.88(1)$ for $N=256, 512$ and $1024$, respectively). 
Configurations were generated using Metropolis Monte Carlo with pivot and local moves.
\footnote{We have verified that the corresponding temperature is indeed close to the finite size $\Theta$-temperature as obtained 
from the inflection point of {$\ree$} as a function of temperature, 
and from the crossing of {$\ree$} as a function of $T$ for two chains with 
a slightly smaller and a slightly larger $N$ ($T=2.90(3)$ for $N=512$).
}

Results for the single chains close to the $\Theta$-point are also displayed in Figs.~\ref{fig:knot_paper}-\ref{fig:knotSizeDistribution}. 
The overall knotting probability (Fig.~\ref{fig:knot_paper}) and the knot spectrum (Fig.~\ref{fig:knotSpectrum}) 
are indeed very similar to the corresponding quantities for 
chains in the melt and differ tremendously from those of a RW. Even the average size of a trefoil 
knot (Fig.~\ref{fig:knotSize31}), as well as the form of the size distribution (Fig.~\ref{fig:knotSizeDistribution}) agrees very well. The latter in particular 
is a strong indication that the structure of chains in a melt is indeed well represented by our $\Theta$-like chains. 
After all, this is not so surprising as long-range bond correlations have been found for both \cite{WiMeBaEA2004prl,ShPaLiRu2008macro}.

\begin{figure}[tb]
\centering
\includegraphics[width=0.98\linewidth]{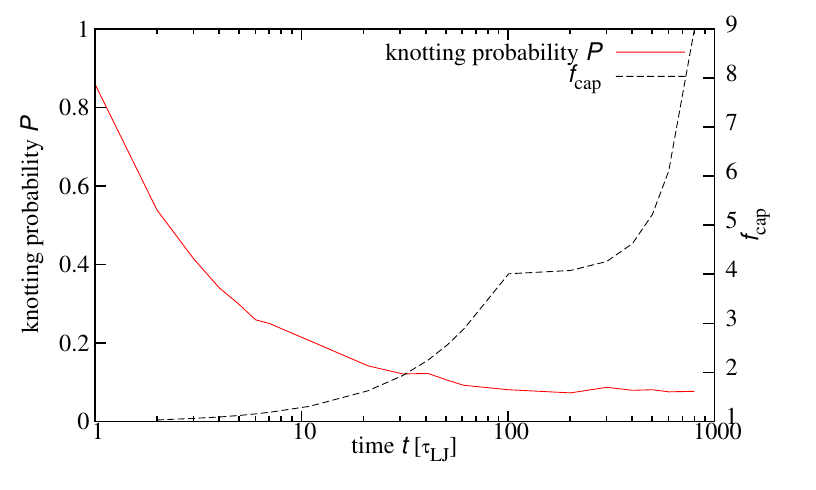}
\caption{Progression of knotting probability during equilibration (continuous line) and force-cap parameter (dashed line, right axis).}
\label{fig:Zeitentwicklung_Knoten}
\end{figure}

Finally, we would like to address an important technical aspect, which arises in the context of equilibration of melts. 
Typically, polymer melts are set up as Gaussian chains \cite{Auhl2003jcp}. 
As explained above, we have increased the strength of the interactions gradually during equilibration as shown in Fig.~\ref{fig:Zeitentwicklung_Knoten}. 
As shown above, knotting behavior without excluded volume interactions 
differs significantly from those in a melt, which could lead to tremendous equilibration times to relax the topology. 
Surprisingly, the setup approach also very quickly equilibrates the topology of the chains in the melt as bonds are still allowed to cross.
In this example the correct knotting probability was obtained after 100 LJ-time units, far before the excluded volume was at full strength.
This underlines that the local crossing and overlap makes an important contribution to the high knotting probability in pure RW,
as seen in the size distribution Fig.~\ref{fig:knotSizeDistribution}.
This is in line with the fact that without rigidity (B0) already moderate excluded volume interactions lead 
to the biggest part of the swelling with respect to a RW \cite{MeWiKrEA2008epje}.
This result also suggests that a topological analysis may in the future 
be carried out by investigating structures emerging from equilibration runs and treating them as independent conformations.

To conclude, the determination of the knot spectrum in polymers is a very fine gauge to measure similarities between structures.
We have used  this tool to test one of the paradigms of polymer physics, namely that a polymer chain in a melt can be described in terms of a RW.
We find that corresponding RWs by far overestimate the occurrence of knots and underestimate their size. 
This finding is attributed to the fact that the local structure of real chains is very different, 
and it is this local structure which accounts for the huge amount of knots in RWs (Fig. \ref{fig:knotSizeDistribution}), 
whereas the mapping to an effective RW is based on the large scale structure.
We found that the knotting properties of chains in the melt are similar to dilute polymer chains close to the $\Theta$-transition, 
which implies that the latter are not well represented by RWs either. This is consistent as corrections to chain ideality have been discovered for both
\cite{WiMeBaEA2004prl,WiBeJoEA2007epl,ShPaLiRu2008macro}.
Although the corrections to ideality have only a minor influence on many measurable chain properties as soon as $C_\infty\gtrsim 3$ \cite{HsKr2016jcp,albert-2dreview}, 
the local remnants of self-avoidance apparently have a strong influence on the knotting probability.

As an outlook, it would be interesting to study the impact of knots on the melt dynamics.
As most knots are rather large, we conjecture that they will be resolved by the regular reptation dynamics, and no additional effects need to be considered in equilibrium.
It would be further interesting to apply the present analysis
to configurations of confined thin polymer films where the global (inter-chain) entanglement density has been shown to decrease \cite{BrRu96macro,MeKrCaWiBa2007epjst,LeDiMeJo2017prl}.
However, it is a debate, to which extent inter-chain entanglements are converted into intra-chain entanglements \cite{SiMaDVBrJo2005prl}.
The tools of the present work could give valuable insights into the intra-chain structure in that case.



\acknowledgments

We thank J. Wittmer and J. Baschnagel for discussions and a critical reading of the manuscript.
We thank GENCI/Equipe@MESO for providing hpc ressources at Universit\'e de Strasbourg and the ZDV Mainz for resources at the Mogon supercomputer.

\bibliography{exportlist2,biblioknot-hm}
\bibliographystyle{apsrev4-1}

\end{document}